\documentclass[12pt,a4paper]{article}
\usepackage{times}
\usepackage{epsfig}
\usepackage{amssymb}
\usepackage{amsmath}
\usepackage{cite}
\usepackage{colordvi}
\usepackage{amsmath}
\usepackage{graphicx}
\usepackage{multicol}
\usepackage{hyperref}

\usepackage{booktabs}
\usepackage{multirow}

\usepackage{longtable}
\usepackage{ltxtable}

\usepackage{appendix}
\usepackage{sectsty}
\sectionfont{\normalsize}
\subsectionfont{\normalsize}

\usepackage{anysize}
\marginsize{2cm}{2cm}{2cm}{1cm}

\def\lapproxeq{\lower .7ex\hbox{$\;\stackrel{\textstyle
<}{\sim}\;$}}
\def\gapproxeq{\lower .7ex\hbox{$\;\stackrel{\textstyle
>}{\sim}\;$}}

\begin {document}


\begin{titlepage}

\vspace{2cm}

\begin{center}
{
NATIONAL CENTRE FOR NUCLEAR RESEARCH, WARSAW \\
}
\vspace{5mm}
{
  July 2, 2012 \\ 
}

\vspace*{2.5cm}

{\LARGE { HEPGEN \\
\vspace*{3mm} 
generator for hard exclusive leptoproduction  }}
\vspace*{1.0cm}

\vspace{5mm}

A.~Sandacz\footnote{E-mail: Andrzej.Sandacz@fuw.edu.pl} and 
P.~Sznajder\footnote{E-mail: Pawel.Sznajder@fuw.edu.pl}

\vspace{5mm}

\begin{small}
 National Centre for Nuclear Research, Warsaw, Poland
\end{small}
\end{center}

\vspace*{2.5cm}

\begin{abstract}
\noindent
HEPGEN is a generator of Monte Carlo events, which is dedicated to studies of hard exclusive
single photon or meson production processes at the COMPASS experiment kinematic domain.
In addition HEPGEN allows to generate also single photon or meson production accompanied by the 
diffractive dissociation of the nucleon, which is one of the main sources of background in 
analyses of exclusive processes.   

\vfill
\end{abstract}
%

\end{titlepage}
{\pagestyle{empty}
~~~~~~~\\
\clearpage
}

\setcounter{page}{3}

\section{Introduction}
 \label{Sec_intro}
HEPGEN (Hard Exclusive Production GENerator) is a dedicated Monte Carlo generator of events, 
which is used for
studies of hard exclusive leptoproduction processes at the COMPASS experiment kinematic domain.  
The studied reactions comprise both single photon production via Deeply Virtual Compton 
Scattering (DVCS) and Bethe-Heitler (BH)
process as well as hard exclusive production of various mesons (HEMP). In addition to
exclusive processes, HEPGEN allows also to generate events of single photon or meson production 
accompanied by the diffractive dissociation of the nucleon, which is one of the main 
sources of background in analyses of exclusive processes.\\

Presently there are four exclusive processes implemented in the generator:\\
\hspace*{5mm}(a) single photon  production (DVCS + BH),\\
\hspace*{5mm}(b) exclusive $\pi ^0$ production,\\ 
\hspace*{5mm}(c) exclusive $\rho ^0$ production,\\
\hspace*{5mm}(d) exclusive $\rho ^+$ production.\\
The implementation of two other exclusive processes, $\phi$ and $J/\psi $ production, 
is under development.

\section{Modelling of exclusive processes}
 \label{Sec_model}
The main features of modelling of the specific processes are shortly described in 
this section.\\

{\bf (a)} The DVCS amplitude is described following the model of Frankfurt, Freund and Strikman (FFS)
~\cite{FFS1,FFS2} with modifications required for COMPASS~\cite{FFS-mod}. The FFS model was 
proposed to describe DVCS data at 
very small $x_{Bj}$ ($\approx 10^{-3})$ measured at HERA 
and it relates the imaginary part of forward DVCS amplitude to the 
structure function $F_2$. In fact the model accounts only for the contribution from 
the GPD $H$,
which is dominant for the unpolarised DVCS cross section at small $x_{Bj}$.  
The modifications and extensions introduced for COMPASS include in particular: 
an additional option for a 
non-factorizable (Reggeized) t-dependence of DVCS amplitude, calculation of the real part of
the amplitude using derivative dispersion relations, and a proper choice of $F_2$ 
parameterisation~\cite{NMC-F2} suitable for the COMPASS kinematic range.

The Bethe-Heitler amplitude is described using formalism of Ref.~\cite{BMK} with 
the replacement of the approximate expressions for the lepton propagators $P_1$ and $P_2$ by
the exact formulae~\cite{guichon}. For the Dirac and Pauli electromagnetic form factors
the dipole formulae are used, which are adequate for the investigated region of 
small $t$ ( $<$ 1 (GeV/$c$)$^2$ ).

Similarly, the DVCS-BH interference term is described following the formalism of 
Ref.~\cite{BMK} with the DVCS amplitude as explained above.\\

{\bf (b)} Exclusive $\pi ^0$ production is simulated using cross sections for longitudinal
($\sigma _L$) 
and transverse ($\sigma _T$) virtual photons according to the GPD model of 
Goloskokov-Kroll~\cite{GKpi0}. In this model the contribution from the transversity
GPD (with helicity flip of exchanged quark) is also included. A
reasonable agreement is achieved with low energy data on exclusive
$\pi ^0$ production from JLAB. 
The generator calculates weights of generated events using a grid of
$\sigma _L$ and $\sigma _T$ cross sections as functions of $W$, $Q^2$ and $t'$.\\ 
   
{\bf (c)} Exclusive $\rho ^0$ production is simulated using the  
shapes of kinematic distributions as a function of $\nu $, $Q^2$ and $p_T^2$, which
were measured by the NMC~\cite{NMC-rho-2} for the deuterium target in a kinematic
range similar to that of COMPASS. The cross section is rescaled using the value from 
the GPD model of Ref.\cite{gk2} at
a selected kinematic point $W_0$ and $Q^2_0$. The ratio $R = \sigma _L/\sigma _T$ is
parameterised using all available measurements from the 
range $5 < W < 300~ {\rm GeV}$.\\

{\bf (d)} Similarly as for $\pi ^0$ meson, exclusive $\rho ^+$ production
is simulated using predictions from the GPD model for this reaction by 
Goloskokov and Kroll~\cite{gk3}.
The grid of $\sigma _L$ and $\sigma _T$ as a function of
$W$, $Q^2$ and $t'$ is used to calculate cross section for each generated event.\\

\section{Diffractive dissociation of the target nucleon}
 \label{Sec_dissociation}   
HEPGEN offers an option to generate also events of leptoproduction of single 
photon or meson
accccompanied by diffractive dissociation of the target nucleon into
several particles, mostly a nucleon and pions.\\

As no experimental data exists for such processes at COMPASS kinematic range, 
a factorisation ansatz is used. We assume that the cross section for such a process
factorises into a part related to the virtual-photon vertex and that for the nucleon
vertex. The former is parameterised using available data or models for
exclusive leptoproduction. For the latter one uses parametrisation of experimental 
data on the proton-proton and proton-deuteron scattering at fixed target Fermilab
experiments, which were carried out with the proton beam energies in the range
100 - 300 GeV~\cite{goulianos}.\\

 The following results from these experiments are used
in HEPGEN.\\
\hspace*{5mm}$\bullet$ The energy dependent ratio of the cross section for single diffractive
dissociation, $p p(d) \rightarrow X p(d)$, to that for elastic scattering,
$p p(d) \rightarrow p p(d)$.\\
\hspace*{5mm}$\bullet$ The shape of the cross section for single diffractive dissociation 
as a function of $M_X$, where $M_X$ is the invariant
mass of the diffractively produced system $X$.\\
\hspace*{5mm}$\bullet$ Charged and neutral hadron multiplicities for the system $X$.\\

\section{HEPGEN output}
 \label{Sec_environment}
The HEPGEN generator runs in a standalone mode producing a file with the output data.
This file can be read-in by COMGEANT, the GEANT-based Monte Carlo package
to simulate responses of COMPASS detectors.
The format of the HEPGEN output data is conformal with the structure of LEPTO/JETSET output.
In particular, it contains kinematic information on all initial, intermediate 
and final particles generated in each event.
Each event is assigned a weight, which is proportional to the cross section
corresponding to the values of kinematic variables for that event.\\

The information from the generator is propagated in its original format through 
the apparatus
simulation and event reconstruction stages, up to the final mini-DST files, which are
used for the physics analyses. Thus this information could be accesses either 
immediately after HEPGEN run, or at the end of the simulation and generation chain.\\

A detailed information on the installation of the HEPGEN package, required input files
and steering parameters data, as well as on how to access the generator output
information is given in the following sections and can be found in Ref.~\cite{hepgen}. 

\section{Technicalities}
\label{Sec_environment}
In this section described are installation and usage of HEPGEN generator.  

\subsection{Installation}
\label{Subsec_installation}
The latest version of HEPGEN source code can be download from the web page given as by Ref. \cite{hepgen}. The source code files are compressed with GNU zip compression utility. Installation is supported by GNU make utility. 

Installation of HEPGEN requires compiler appropriate for FORTRAN 77 standard, e.g. g77. In addition, installation requires CERNLIB library to be installed. After standard compilation the binary executable is created, {\tt{hepgen.exe}}, which loads libraries dynamically. 

\subsection{Input and output files}
\label{Subsec_io}
HEPGEN use three input files~ - ~{\tt{hepgen.data}}, ~{\tt{fort.9}} and {\tt{fort.31}}. The text file {\tt{hepgen.data}} is the configuration file of HEPGEN. It contains steering data, like for instance selection of simulated process, or used range of kinematic variables. All available options are described in Sec. \ref{Subsec_steering}. Example of {\tt hepgen.dat} file is given in \ref{App_conf_file}. The text file {\tt{fort.9}} contains so-called random seed, i.e. integer number used for an initialization of a random number generator. The file is updated by the generator after finishing the job. The binary file {\tt{fort.31}}, referred to as the beam file in the following, contains a set of beam particles. Each entry in the beam file corresponds to a single beam particle. An entry contains information about direction at a given space point, momentum and a type of beam particle which can be either true beam particle or beam halo particle. Simulation can be done either for true beam particles or for halo particles, or for both types (cf. Sec. \ref{Subsec_steering}). The beam file is read by FORTRAN unformatted I/O stream mechanism. Instruction how to create a new beam file can be found in {\tt{create\_beam.F}} file, which is enclosed to the HEPGEN source code. 

HEPGEN output files are {\tt{fort.11}} and {\tt{fort.40}}. The binary file {\tt{fort.11}} contains output data. Access to this data is described in Sec. \ref{Subsec_access}. The binary file {\tt{fort.40}} contains histograms in HBOOK format, e.g. distributions of generated kinematics variables.  

The log of HEPGEN goes to the standard output. It contains for instance values of steering parameters and JETSET output of the first 30 events, which is useful for checks.  

\subsection{Steering parameters}
\label{Subsec_steering}
All steering parameters are summarized in Tab. \ref{tab:Subsec_steering_0}. 
\newpage 
{\small
\begin{center}
\begin{longtable}{l p{0.128\textwidth} p{0.15\textwidth} p{0.55\textwidth}}
\caption{\normalsize{Description of steering parameters of HEPGEN. For each parameter the type ($\mathrm{int}$/$\mathrm{real}$) of argument (or arguments) is given and a list of possible values for those arguments which can take only specific, predefined values.}}
\label{tab:Subsec_steering_0}

\endfirsthead
\endhead
\endfoot
\endlastfoot

\toprule
\multirow{2}{*}{Name} & Type of\phantom{xxxx} argument(s)& \multirow{2}{*}{Possible values} & \multirow{2}{*}{Description} \\
\midrule
NGEV     & $\mathrm{int}$                 &           & number of events to be generated \\
PROC     & $\mathrm{int}$                 &           & type of process to be generated \\ 
         &                    &  \centering{5}        & hard single production mode \\
VMES     & $\mathrm{int}$ $\mathrm{int}$            &           & codes of generated particle and its decay mode\\
         &                    &  \centering{\phantom{xx}0 ~~~ 0\phantom{xx}}      & $\gamma$ ~~~~~~(DVCS and BH) \\
         &                    &  \centering{\phantom{xx}1 ~~~ 22\phantom{x}}     & $\pi^0 \rightarrow \gamma + \gamma$ \\
         &                    &  \centering{\phantom{xx}2 ~~~ 211}    & $\rho^0 \rightarrow \pi^{+} + \pi^{-}$ \\
         &                    &  \centering{\phantom{xx}3 ~~~ 321}    & $\phi \rightarrow K^{+} + K^{-}$ ~~~~~~~(\emph{under development}) \\
         &                    &  \centering{\phantom{xx}4 ~~~ -13}    & $\mathrm{J}/\Psi \rightarrow \mu^{+} + \mu^{-}$ ~~~~~(\emph{under development}) \\
         &                    &  \centering{\phantom{xx}5 ~~~ -11}    & $\mathrm{J}/\Psi \rightarrow e^{+} + e^{-}$ ~~~~~~(\emph{under development}) \\
         &                    &  \centering{\phantom{xx}7 ~~~ 211}    & $\rho^+ \rightarrow \pi^{+} + \pi^{0}$ \\
DIFF     & $\mathrm{int}$                 &           & if set, events with nucleon diffractive dissociation are also generated \\
         &                    &  \centering{0}        & diffractive dissociation off (only exclusive events) \\
         &                    &  \centering{1}        & diffractive dissociation on (both exclusive and diffractive dissociation events) \\
BMRD     & $\mathrm{int}$                 &           & type of beam particles \\
         &                    &  \centering{0}        & beam file not used, direction of the beam is parallel to $z$ axis and momentum of 
                                             the beam is given by the first argument of BEAM \\
         &                    &  \centering{1}        & true beam only \\
         &                    &  \centering{2}        & beam halo only \\
         &                    &  \centering{3}        & true beam and beam halo \\
BEAM     & $\mathrm{real}$ $\mathrm{real}$ $\mathrm{int}$ $\mathrm{int}$ $\mathrm{int}$     & & definition of colliding particles, first and second arguments refer to  
                                       the beam and the target energies, the third one is a PID of the beam particle, 
                                       two last refer to the mass number and atomic number of the target, respectively \\
BPAR     & $\mathrm{real}$ $\mathrm{real}$     & & lepton beam charge and polarisation \\
MACC     & $\mathrm{real}$                     & & azimuthal acceptance of scattered muons (\emph{this option is obsolete}) \\
TPAR     & $\mathrm{real}$ $\mathrm{real}$ $\mathrm{real}$ $\mathrm{real}$& & target parameters, the first argument is an average value of mass number of the target nuclei, 
                                       second is a fraction of coherent events to be generated in the sample, two last arguments are 
                                       values of slopes of $t'$ distributions for coherent and incoherent events, respectively \\
ALFA     & $\mathrm{real}$                     & $\alpha$ & dependence of cross section on the mass number of target nuclei, $\alpha$ is defined as 
                                       ${\widetilde \sigma_{A}} = \sigma_{p} \cdot A^{\alpha - 1}$, where $\widetilde \sigma_{A}$ is the cross section per 
                                       nucleon for the reaction on the nucleus of mass number $A$, $\sigma_{p}$ is the cross section 
                                       for the reaction on the free proton \\
CUTL     & $\mathrm{real}$ $\mathrm{real}$ $\mathrm{real}$ $\mathrm{real}$ $\mathrm{real}$ $\mathrm{real}$ $\mathrm{real}$ $\mathrm{real}$ $\mathrm{real}$ $\mathrm{real}$ &  & minimum and maximum 
                                       values of Bjorken scaling variable ($x_{Bj}$), fraction of the 
                                       lepton energy lost in LAB ($y$), negative four-momentum squared of the virtual photon ($Q^{2}$), 
                                       total energy squared in the virtual photon - nucleon system ($W^{2}$), energy of the virtual photon in LAB ($\nu$), 
                                       energy of scattered beam particle and azimuthal angle of scattered beam particle respectively 
                                       (\emph{most of those parameters became obsolete, the active ones are those related to $Q^2$ and $\nu$ variables}) \\
TLIM     & $\mathrm{real}$ $\mathrm{real}$                 & &  minimum and maximum values of $t'$ variable, defined as $t'=|t|-t_{0}$, where $t$ is four-momentum transfer to the 
                                       target and $t_{0}$ is minimal value of $|t|$ allowed by the kinematics of the event \\ 
REGG     & $\mathrm{real}$ $\mathrm{real}$ $\mathrm{real}$ & \centering{$b_{0}$ $b_{1}$ $x_{0}$} & parametrization of slope $b$ of $t$-distribution as a function of $x_{Bj}$, 
                                       parametrization is given by $b = b_{0} + 2 \cdot b_{1}\log(x_{0}/x_{Bj})$ (\emph{these parameters are used only for DVCS production}) \\ 
\bottomrule
\end{longtable}
\end{center}
}

\subsection{Access to output data}
\label{Subsec_access}
The output data from HEPGEN generator are stored in {\tt{fort.11}} file. The file can be used either by COMGEANT or directly by the user. In the first case, informations from the generator are available in PHAST, which is the framework for the COMPASS data analysis. Details how to access these informations in PHAST are given in Ref. \cite{hepgen}. In the case of direct access the user should use the interface subroutine {\tt{read\_interface.F}}, which is enclosed to the HEPGEN source code. Information given below is relevant only when direct access to HEPGEN output data is needed (not needed for PHAST users).

The output file {\tt{fort.11}} contains header record and event records. Each event record contains data, which are specific for corresponding event. The data, both from header and event records, are stored by {\tt{read\_interface.F}} in the following FORTRAN common blocks 
\begin{center}
\parbox{0.7\textwidth}{ 
{\tt
\rule{0px}{12px}LEPTOU, USERVAR, LUJETS,
}
}
\end{center}
\rule{0px}{12px}where {\tt LEPTOU, USERVAR, LUJETS} are common blocks, defined as
\begin{center}
\parbox{0.7\textwidth}{ 
{\tt
    \rule{0px}{12px}REAL CUT(14),PARL(30),X,Y,W2,Q2,U \\
    INTEGER LST(40) \\
    COMMON /LEPTOU/ CUT,LST,PARL,X,Y,W2,Q2,U
}
}
\end{center}
\begin{center}
\parbox{0.7\textwidth}{ 
{\tt
   \rule{0px}{15px}REAL USERVAR(20) \\
   COMMON /USERVAR/ USERVAR 
}
}
\end{center}
\begin{center}
\parbox{0.7\textwidth}{ 
{\tt
   \rule{0px}{15px}INTEGER*4 N,K(4000,5) \\
   REAL*4 P(4000,5),V(4000,5) \\
   COMMON/LUJETS/N,K,P,V 
}
}
\end{center}
\rule{0px}{12px}Description of the most useful variables is given in Tab. \ref{tab:Subsec_access_0}. Note, that the format of common block {\tt LUJETS} and the values of elements of {\tt K} array are consistent with those used in JETSET. Detailed description of JETSET can be found in Ref. \cite{PIDcodes}.

\newpage 

\begin{table}[!ht]
\caption{\normalsize{Description of the most useful variables of HEPGEN output.}} 
\label{tab:Subsec_access_0}
{\small
\begin{center}
\begin{tabular}{l l p{0.15\textwidth} p{0.47\textwidth}}

\toprule
Variable name     & &\centering{Type}& Description \\
\midrule
USERVAR(3) & &    \centering{$\mathrm{real}$}   & weight of event\\
           & &                      & for exclusive photon production weight corresponds to $|\mathrm{DVCS}|^{2} + |\mathrm{BH}|^{2} + \mathrm{Int}$ \\
USERVAR(16) & &    \centering{$\mathrm{real}$}  & for exclusive photon production weight for $|\mathrm{DVCS}|^2$ \\
USERVAR(17) & &    \centering{$\mathrm{real}$}  & for exclusive photon production weight for $|\mathrm{BH}|^2$ \\
LST(24) &  & \centering{$\mathrm{int}$} &  yes ($1$) if diffractive dissociation event, no ($0$) if exclusive event \\
X & & \centering{$\mathrm{real}$}  & value of Bjorken scaling variable ($x_{Bj}$)\\
Y & & \centering{$\mathrm{real}$}  & value of lepton energy lost in LAB ($y$) \\
W2 & & \centering{$\mathrm{real}$}  & value of total energy squared in the virtual photon - nucleon system ($W^{2}$) \\
Q2 & & \centering{$\mathrm{real}$}  & value of negative four-momentum squared of the virtual photon ($Q^{2}$)\\
U & & \centering{$\mathrm{real}$}  & value of  energy of the virtual photon in LAB ($\nu$)\\
N & &                          \centering{$\mathrm{int}$}    &  number of initial, intermediate and final particles\\
K($I$, 1) & $I$ = \phantom{x}1, ... , N & \centering{$\mathrm{real}$} & status code of $I$-th particle \\
K($I$, 2) & $I$ = \phantom{x}1, ... , N & \centering{$\mathrm{real}$} & PID of $I$-th particle, possible codes are summarized in Ref. \cite{PIDcodes} \\
K($I$, 3) & $I$ = \phantom{x}1, ... , N & \centering{$\mathrm{real}$} & index of parent particle of particle $I$ \\
K($I$, 4) & $I$ = \phantom{x}1, ... , N & \centering{$\mathrm{real}$} & index of first daughter of particle $I$ \\
K($I$, 5) & $I$ = \phantom{x}1, ... , N & \centering{$\mathrm{real}$} & index of last daughter of particle $I$ \\
P($I$, 1) & $I$ = \phantom{x}1, ... , N & \centering{$\mathrm{real}$} & momentum in $x$ direction of particle $I$\\
P($I$, 2) & $I$ = \phantom{x}1, ... , N & \centering{$\mathrm{real}$} & momentum in $y$ direction of particle $I$\\
P($I$, 3) & $I$ = \phantom{x}1, ... , N & \centering{$\mathrm{real}$} & momentum in $z$ direction of particle $I$\\
P($I$, 4) & $I$ = \phantom{x}1, ... , N & \centering{$\mathrm{real}$} & energy of particle $I$\\
P($I$, 5) & $I$ = \phantom{x}1, ... , N & \centering{$\mathrm{real}$} & mass of particle $I$\\

\bottomrule
\end{tabular}
\end{center}
}
\end{table}

\appendix
\newpage 

\section{Example of {\tt hepgen.data} file}
\label{App_conf_file}

\begin{table}[!ht]
\label{tab:App_conf_file_0}
{\small
{\tt
\begin{center}
\begin{tabular}{l p{0.08\textwidth} p{0.08\textwidth} p{0.08\textwidth} p{0.08\textwidth} p{0.08\textwidth} p{0.08\textwidth} p{0.08\textwidth} p{0.08\textwidth}}
LIST  &&&&&&&& \\
      &&&&&&&& \\
      \multicolumn{9}{l}{\it * number of events} \\
NGEV  &  10000 &&&&&&&\\
      \multicolumn{9}{l}{\it * physics process} \\
PROC  &  5 &&&&&&&\\
      \multicolumn{9}{l}{\it * produced particle and its decay mode} \\
VMES  &  0 &  0 &&&&&&\\ 
      \multicolumn{9}{l}{\it * diffractive dissociation switch} \\
DIFF  &  1 &&&&&&&\\
      \multicolumn{9}{l}{\it * beam file} \\
BMRD  &  1 &&&&&&&\\
      \multicolumn{9}{l}{\it * colliding particles} \\
BEAM  &  160.0& 0.93827& ~-13& 1& 1&&& \\
      \multicolumn{9}{l}{\it * beam charge and polarisation} \\
BPAR  &  1.0& -1.0&&&&&& \\
      \multicolumn{9}{l}{\it * acceptance of scattered muon} \\
MACC  &  0.050&&&&&&& \\
      \multicolumn{9}{l}{\it * target parameters} \\
TPAR  &  1.0& -0.1& 52.2& 5.0&&&& \\
      \multicolumn{9}{l}{\it * dependence of cross sections on mass number $A$} \\
ALFA  &  0.75&&&&&&& \\
      \multicolumn{9}{l}{\it * limits for $x_{Bj}$, $y$, $Q^2$, $W^2$, $\nu$, $E'$, $\phi$ variables} \\
CUTL  &  0.0001& 1.0& 0.0& 1.0& 0.5& 80.0& 0.00& 1000.0 \\
      &  5.0& 155.0& 0.0& 200.0& 0.00& 6.28318&& \\
      \multicolumn{9}{l}{\it * limits for $t'$ variable} \\
TLIM  &  0.005& 1.0&&&&&& \\
      \multicolumn{9}{l}{\it * parameterisaton of $x_{Bj}$, $t$ correlation} \\
REGG  &  4.94116& ~0.042& 0.8&&&&& \\
      &&&&&&&& \\
END   &&&&&&&& \\
\end{tabular}
\end{center}
}
}
\end{table}


\noindent
\begin{thebibliography}{99}

\bibitem{FFS1} L.L.~Frankfurt, A.~Freund, M.~Strikman, Phys. Rev. {\bf D58} (1998) 114001;\\Erratum-ibid, {\bf D59} (1999) 119901.

\bibitem{FFS2} L.L.~Frankfurt, A.~Freund, M.~Strikman, Phys. Lett. {\bf B460} (1999) 417.

\bibitem{FFS-mod} A.~Sandacz, {\it Modifications to FFS model and predictions},
 April 22, 2009,\\
http://wwwcompass.cern.ch/compass/gpd/index.html

\bibitem{NMC-F2} NMC Collaboration, M.~Arneodo {\it et al.}, Phys. Lett. {\bf B364} (1995) 107.

\bibitem{BMK} A.V.~Belitsky, D.~M\"{u}ller, A.~Kirchner, Nucl. Phys. {\bf B629} (2002) 323.

\bibitem{guichon} P.A.M.~Guichon, private communication.

\bibitem{GKpi0} S.V.~Goloskokov and P.~Kroll, arXiv: 1106.4897 [hep-ph].

\bibitem{NMC-rho-2} NMC Collaboration, M.~Arneodo {\it et al.}, Nucl. Phys. {\bf B429} (1994) 503.

\bibitem{gk2} S.V.~Goloskokov and P.~Kroll, Eur. Phys. J. {\bf C53} (2008) 367.

\bibitem{gk3} S.V.~Goloskokov and P.~Kroll, Eur. Phys. J. {\bf C59} (2009) 809.



\bibitem{goulianos} K.~Goulianos, Phys. Rep. {\bf 101}, No. 3 (1983) 169.

\bibitem{hepgen} A.~Sandacz and P.~Sznajder, http://project-gpd-full-chain-mc.web.cern.ch\\/project-gpd-full-chain-mc/hepgen/index.html.

\bibitem{PIDcodes} T.~Sjostrand, hep-ph/9508391

\end{thebibliography}
\end{document}